\documentclass[a5paper,10pt]{article}
\usepackage[pass]{geometry}
\usepackage{graphicx}
\usepackage{amsmath}
\usepackage{cite}

\begin{document}

\pagestyle{plain}

\begin{center}
    {\bf Superheat conductivity and electric activity of superfluid systems\\
    \vspace*{10pt}
    {S. I. Shevchenko}, {A. M. Konstantinov}}\\
    \vspace*{10pt}
    \textit{B. Verkin Institute for Low Temperature Physics and Engineering,\\Kharkov 61103, Ukraine\\
    E-mail: shevchenko@ilt.kharkov.ua}
\end{center}

\vspace*{10pt}

\begin{abstract}
In this article it is shown that in He\nobreakdash-II placed in a magnetic field $\mathbf{H}$, the heat flow under the influence of a temperature gradient $\nabla T$ causes appearance of an electric field $\mathbf{E}\sim\mathbf{H}\times\nabla T$. This effect occurs in superfluid dielectric systems due to their specific property named superheat conductivity. The magnitude of the electric field significantly depends on the shape of the helium sample and the magnetic field direction relative to the sample. The effect exists both at static and non-stationary temperature gradient (e. g. during propagation of second sound).
\end{abstract}

Recently, the problem of the electrical activity of superfluid systems has attracted considerable interest. The appearance of this problem is connected with the experimental work \cite{bib:R, bib:RR}, performed at ILTPE in Kharkiv about fifteen years ago. In the works \cite{bib:R} it was found, in particular, that the standing wave of  second sound in the resonator led to the appearance of a potential difference $\Delta\varphi$ between the ends of the resonator proportional to the temperature difference $\Delta T$ between these ends. Authors of \cite{bib:RR} detected a mechano-electric effect: torsional vibrations of a cell with a torsion oscillator were accompanied by oscillations of the potential difference between the cell surface and the electrode on its axis. The results of \cite{bib:R} have been confirmed in several recent experiments \cite{bib:Cha1,bib:Cha2,bib:Cha3,bib:Jap}.

The detected effects became the object of numerous theoretical studies. Polarization of dielectric systems by acceleration \cite{bib:M} was predicted (a phenomenon similar to the Stewart-Tolman effect in metals). It has been established that a quantized vortex creates an axially symmetric polarization around itself ("polarization hedgehog") \cite{bib:N,bib:SR}, and the flow of the superfluid component relative to the vortex leads to the appearance of a dipole moment in the vortex \cite{bib:SK1}. The polarization vector due to the dipole moments of vortex rings in the field of Van der Waals forces \cite{bib:AN1,bib:AN2} has been calculated. A model is proposed to explain the results of experiments by the appearance of dipole moments in helium atoms near the walls of the vessel with He\nobreakdash-II \cite{bib:SRJetpLett} and near the measuring electrodes \cite{bib:G,bib:Tom}. However, the generally accepted point of view on the nature of the phenomena observed in \cite{bib:R,bib:RR,bib:Cha1,bib:Cha2, bib:Cha3,bib:Jap} is currently absent.

The role of near-surface polarized layers is especially significant in thin superfluid films. In such films, in the presence of a periodically varying temperature difference at the edges of the film, a third sound wave propagates over the film. The wave is accompanied by variations in the film thickness, which lead to oscillations of the dipole moment induced by the wall and the appearance of an oscillating electric field in the surrounding space. This effect is predicted in \cite{bib:SKJetpLett, bib:SK2}.

In this paper, attention is drawn to the mechanism of thermal polarization that is possible in superfluid dielectric systems in the presence of a magnetic field. He\nobreakdash-II is known to be characterized by two unique properties, one of which is superfluidity, and the other is often called superheat conductivity (this name is suggested in \cite{bib:Keesom}). The tremendous value of the thermal conductivity in the superfluid state is due to the fact that the heat transfer in He\nobreakdash-II is associated with the movement of the normal component. In this case, the mass flow carried by the normal component is compensated by the flow carried by the superfluid component. An extremely important circumstance is the fact that the condition for the absence of mass flow should be satisfied only on average, i.e. the mass flux averaged over the system area should be zero. In general, the local mass flow should not vanish.
Since the movement of any dielectric in a magnetic field leads to its polarization, in the presence of a magnetic field a local dipole moment in superfluid helium will be associated with a local mass flow generated by a temperature gradient and an average electric field may occur outside the system.
We can say that the effect under discussion is analogous to the Nernst-Ettingshausen effect in conductors. But it turned out that in superfluid systems the magnitude of the emerging electric potential is very sensitive to the geometry of the system and the orientation of the magnetic field. Thus, there is no effect in a circular capillary, and in the case of an elliptical capillary with $ a $ and $ b $ semi-axes, the electric field at $ a \gg b $ changes by more than three orders of magnitude depending on whether the magnetic field is directed along the large or the small axis. This article is devoted to determining the nature of the dependence of the electric field on the magnetic field magnitude, its direction and on the observation point.

Let us proceed from the expression obtained by Minkowski (see, for example, \cite {bib:LL8}) for the induction $ \mathbf {D} $ in a dielectric with a dielectric constant $ \epsilon $ and magnetic permeability $\mu $ as it moves at a speed of $ \mathbf {v} $
\begin {align}
\mathbf {D} = \epsilon \mathbf {E} + \frac {\epsilon \mu -1} {c} \left (\mathbf {v} \times \mathbf {H} \right). \label {eq:1}
\end {align}
Here $ \mathbf {E} $ and $ \mathbf {H} $ are the strengths of the stationary electric and magnetic fields. The expression (\ref {eq:1}) is valid with linear accuracy in $ v / c $.

In the case of $^4$He,  with a reasonable degree of accuracy, we can assume that $ \mu = 1 $. The relationship of the dielectric constant $ \epsilon $ to the polarizability $ \alpha $ of a $ ^4 $He atom and the density of atoms $ n $ can be found using the Clausius-Mossotti relation
\begin {align}
\frac {\epsilon - 1} {\epsilon + 2} = \frac {4 \pi n \alpha} {3}, \label {eq:2}
\end {align}
which is a consequence of the fact that an atom is polarized not by the mean field $ \mathbf {E} $, but by a local electric field. For $ ^4 $He the density is $ n = 2 \cdot 10 ^ {22} $ cm$ ^ {- 3} $, the polarizability $ \alpha = 2 \cdot 10 ^ {- 25} $ cm$ ^3 $ and the product $ n \alpha $ is substantially less than one. Therefore, $ \epsilon - 1 $ is an expansion in powers of $ n \alpha $, in which it is sufficient to keep only the first term of the expansion $ 4 \pi n \alpha $.

Taking into consideration that the induction is $ \mathbf {D} = \mathbf {E} +4 \pi \mathbf {P} $ and that the mass flux is $ \mathbf {j} = Mn \mathbf {v} $ ($ M $ - mass of the helium atom), we obtain from (\ref {eq:1}) in the absence of an electric field
\begin {align}
\mathbf {P} = \frac {\alpha} {Mc} \left [\mathbf {j} \times \mathbf {H} \right]. \label {eq:3}
\end {align}
In the general case, in superfluid helium, $ \mathbf {j} = {{\rho} _ {n}} {{\mathbf {v}} _ {n}} + {{\rho} _ {s}} {{ \mathbf {v}} _ {s}} $, where $ {{\mathbf {v}} _ {n}} $, $ {{\mathbf {v}} _ {s}} $ are the velocities of normal and superfluid component, and $ {{\rho} _ {n}} $, $ \rho_s $ are their mass densities. The expression (\ref {eq:3}) does not depend on the method of setting He\nobreakdash-II in motion. Below, we will assume that the motion of helium is induced by a temperature gradient.

Let us consider the problem of stationary heat flow in He\nobreakdash-II which fills a capillary, between whose ends a small temperature difference $ \Delta T $ is maintained. With laminar motion (analysis of which we restrict ourselves to), the speeds $ {{\mathbf {v}} _ {n}} $ and $ {{\mathbf {v}} _ {s}} $ do not depend on the longitudinal coordinate (along the capillary) and the equations of two-fluid hydrodynamics have the form
\begin {align}
{{\eta} _ {n}} {{\nabla} ^ {2}} {{\mathbf {v}} _ {n}} = \frac {{{\rho} _ {n}}} {\rho} \nabla P + {{\rho} _ {s}} S \nabla T, \label {eq:4} \\
\nabla P = \rho S \nabla T. \label {eq:5}
\end {align}
Here $ {{\eta} _ {n}} $ is the viscosity coefficient of the normal component, $ \rho = {{\rho} _ {n}} + {{\rho} _ {s}} $ is the total mass density, $ \nabla P $ and $ \nabla T $ are pressure and temperature gradients, $ S $ is specific entropy. The second of these equations, called the London equation, is a consequence of the mechanical equilibrium that occurs when a superfluid flows between the ends of the capillary. From (\ref {eq:4}) and (\ref {eq:5}) the following equation is obtained:
\begin {align}
{{\eta} _ {n}} {{\nabla} ^ {2}} {{\mathbf {v}} _ {n}} = \nabla P, \label {eq:6}
\end {align}
which is equivalent to the Poiseuille equation in classical hydrodynamics.

The superfluid motion velocity cannot depend on the transverse coordinate, since $ \nabla \times {{\mathbf {v}} _ {s}} = 0 $. Its value can be found from the condition of the absence of the total mass flow, i.e. from the condition $ {{\rho} _ {n}} \left \langle {{\mathbf {v}} _ {n}} \right \rangle + {{\rho} _ {s}} {{\mathbf { v}} _ {s}} = 0 $. Here $ \left \langle {{\mathbf {v}} _ {n}} \right \rangle $ is the velocity of the normal component, averaged over the cross-section area of the capillary. Given this condition, we obtain from (\ref {eq:3})
\begin {align}
\mathbf {P} = \frac {\alpha {{\rho} _ {n}}} {Mc} \left [\left ({{\mathbf {v}} _ {n}} - \left \langle { {\mathbf {v}} _ {n}} \right \rangle \right) \times \mathbf {H} \right]. \label {eq:8}
\end {align}
We see that in the presence of a magnetic field, the polarization is locally non-zero, but the total dipole moment vanishes along with the total mass flow. The electrical potential outside the system is determined by the expression
\begin{align}
\varphi ({{\mathbf{r}}_{0}})=\int{\,\frac{\mathbf{P}\cdot \left( {{\mathbf{r}}_{0}}-\mathbf{r} \right)}{{{\left| {{\mathbf{r}}_{0}}-\mathbf{r} \right|}^{3}}}{{d}^{3}}r},\label{eq:9}
\end{align}
where $ {{\mathbf {r}} _ {0}} = ({{r} _ {0}}, {{\theta} _ {0}}, {{z} _ {0}}) $ is the radius vector of the observation point. Therefore, vanishing of the total dipole moment in the general case does not lead to the absence of electric fields outside the system.

The solution of (\ref {eq:6}) depends on the geometry of the problem. It seems natural to choose a circular cross-section cylindrical tube as a capillary. However, the calculation shows that in this case the potential outside the capillary is identically zero. This result is a consequence of high symmetry of such a system. Consideration of nonlinear in $ n \alpha $ terms in the expansion of the expression for $ \epsilon $ does not change the symmetry of the problem, therefore, an axially symmetric flow along a cylindrical capillary is not accompanied by the appearance of an electric field outside the capillary even if the nonlinear dependence $ \epsilon (n) $ is taken into account. The potential $ \varphi ({{\mathbf {r}} _ {0}}) $ is nonzero if we switch from a circular capillary to a capillary, for example, with an elliptical cross section.

Consider the case of an elliptical capillary with a small temperature difference $ \Delta T $ between its ends. Let the semi-axes of the ellipse $ a $ and $ b $ lie along the axes $ \mathbf {\hat {x}} $ and $ \mathbf {\hat {y}} $, respectively. The expression for the normal component velocity can be found in \cite {bib:LL6}. Substituting this velocity of the normal component into (\ref {eq:8}) and assuming that the magnetic field $ \mathbf {H} $ is directed along the $ \mathbf {\hat {y}} $ axis, we find the polarization
\begin {align}
\mathbf {P} (x, y) = - {{P} _ {0}} \frac {{{a} ^ {2}}} {{{a} ^ {2}} + {{b} ^ {2}}} \left (\frac {1} {2} - \frac {{{x} ^ {2}}} {{{a} ^ {2}}} - \frac {{{y} ^ {2}}} {{{b} ^ {2}}} \right) \mathbf {\hat {x}}. \label {eq:14}
\end {align}
Here we use the notion
\begin {align}
{{P} _ {0}} = \frac {\alpha {{\rho} _ {n}} H} {Mc} \frac {\rho S \Delta T} {2 {{\eta} _ {n }} L} {{b} ^ {2}}, \label {eq:15}
\end {align}
where $L$ is the capillary length.

The electric potential outside the fluid can be obtained by substituting (\ref{eq:14}) into (\ref{eq:9}). However, integration cannot be performed for arbitrary coordinates of the observation point $ {{x} _ {0}} $ and $ {{y} _ {0}} $. An analytical expression can be obtained for $ {{y} _ {0}} = 0 $. But even in this case, the answer is very cumbersome. We give the answer only for two special cases. We introduce the notation $ \gamma = b / a $. Then the potential on the capillary surface at $ x_0 = a $ is
\begin{align}
\varphi =\frac{2\pi b{{P}_{0}}}{3}\frac{1-\gamma }{\left( 1+{{\gamma }^{2}} \right){{\left( 1+\gamma  \right)}^{2}}}.\label{eq:18}
\end{align}
At $ \gamma = 1 $, i.e. at $ a = b $, when the ellipse turns into a circle, the expression (\ref {eq:18}) vanishes in accordance with the statement made above.

At $ \gamma \ll 1 $ (in this case, the system simulates a slit, with a magnetic field applied across the slit)
\begin{align}
\varphi ({{x}_{0}},0)=\frac{2\pi b{{P}_{0}}}{3}\left\{ \sqrt{{{x}_{0}}^{2}-1} \right. + \left. {{x}_{0}}\left[ 4{{x}_{0}}\left( {{x}_{0}}-\sqrt{{{x}_{0}}^{2}-1} \right)-3 \right] \right\}.\label{eq:20}
\end{align}
Here $ {{x} _ {0}} $ is measured in units of $ a $.

With $ \gamma \gg1 $, the system again models a slit, but the magnetic field is applied along the slit. In this case, the potential is several orders of magnitude smaller than with $ \gamma \sim1 $ and $ \gamma \ll1 $.

For arbitrary $ {{x} _ {0}} $ and $ {{y} _ {0}} $, numerical integration allows finding potential $ \varphi $ when moving from points with coordinates $ ({{x} _ {0} }, {{y} _ {0}} = 0) $ to points $ ({{x} _ {0}} = 0, {{y} _ {0}}) $. One would expect a monotonic decrease in the potential $ \varphi ({{x} _ {0}}, {{y} _ {0}}) $. It turns out that this is not the case. Below are graphs of the dependences of the potential $ \varphi $ at the capillary surface on the polar angle $ \theta = \arctan ({{y} _ {0}} / {{x} _ {0}}) $ for $ \gamma \sim1 $ and $ \gamma \ll1 $.

\begin{figure}[h]
\begin{center}
\includegraphics{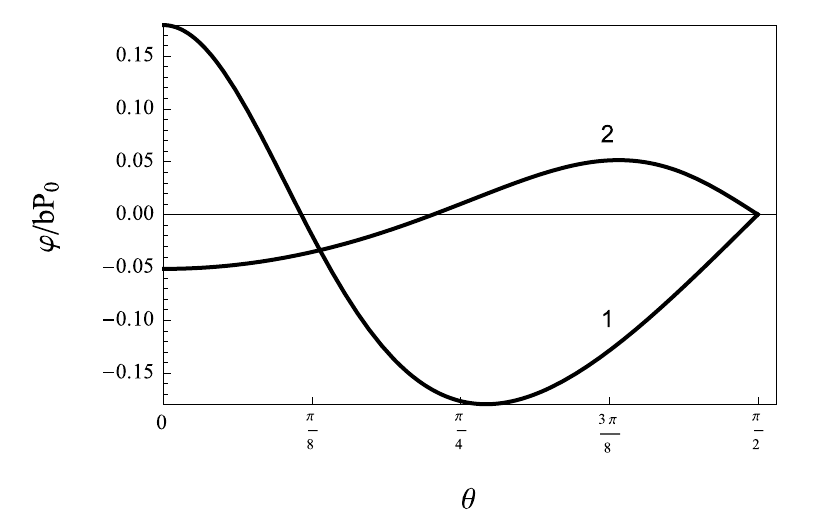}
\end{center}
\caption{Angular dependencies of the potential $ \varphi $ on the capillary surface at $ \gamma \sim 1 $. Curve 1 corresponds to the magnetic field directed along the minor semiaxis of the ellipse, curve 2 corresponds to the field along the semimajor axis of the ellipse}
\end{figure}

\begin{figure}[h]
\begin{center}
\includegraphics{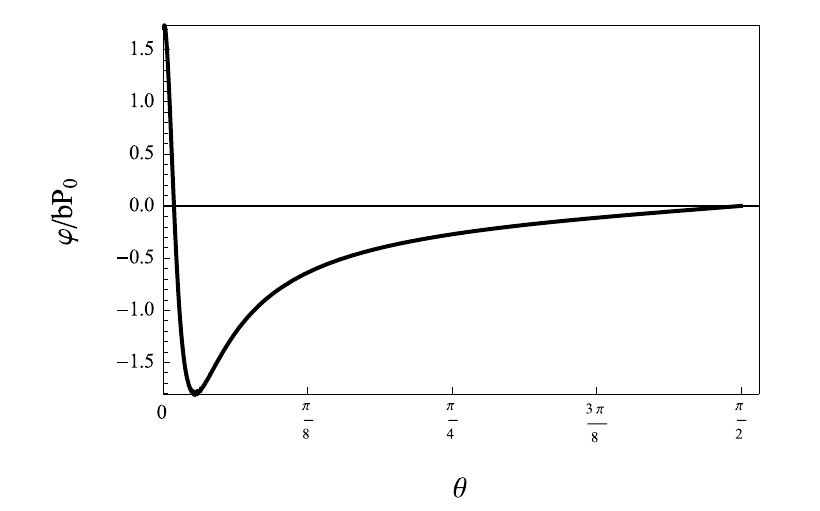}
\end{center}
\caption{Angular dependence of the potential $ \varphi $ on the capillary surface at $ \gamma = 0.05 $}
\end{figure}

The analysis shows that the most favorable case for obtaining a large potential $ \varphi $ is the case of $ \gamma \ll1 $ (a magnetic field is applied across the slit). The maximum potential is reached at small angles $ \theta $. In particular, when $ \theta = 0 $ and $ x_0 = 1 $ the potential is
\begin {align}
\varphi = \frac {\pi {{b} ^ {3}}} {3} \frac {\alpha {{\rho} _ {n}} H} {Mc} \frac {\rho S \Delta T } {{{\eta} _ {n}} L}. \label {eq:22}
\end {align}

The dependence of the potential $ \varphi $ on temperature is determined by the temperature dependence of the normal density $ \rho_n $, viscosity $ \eta_n $, and also the specific entropy $ S $. The dependence on temperature of the total density $ \rho $ and polarizability $ \alpha $ can be neglected. It seems that the potential value $ \varphi $ can be significantly changed by changing the value of $ b $. However, such a change is possible only within certain limits. The reason is that with an increase in $ b $, the laminar motion of a fluid transforms into a turbulent one.

In the pioneering work \cite{bib:Vin1} it was shown that when the heat flux  $ W = {{\rho} _ {s}} ST \left| {{\mathbf {v}} _ {n}} - {{\mathbf {v}} _ {s}} \right| $ reaches the critical value $ {{W} _ {c1}} $, the laminar motion is destroyed by appearance in superfluid liquid of quantized vortices and vortex rings that form a vortex tangle (T-1 state). Further studies showed that there is a second critical heat flux $ {{W} _ {c2}} $, above which the vortex density increases significantly (T-2 state). The transition from the T-1 state to the T-2 state is associated with the transition to the turbulent state of the normal component (see, for example, \cite{bib:TurbPRL, bib:Tsu}). The result greatly depends on the shape of the channel. Two transitions take place in elliptic channels with dimensions $ a \approx b $. When $ a \gg b $ there is only one transition to the T-2 state \cite{bib:Tough}.

In the T-1 state, rings with a radius of the order of the size of the capillary $ b $ are critical. For the appearance of such rings, the difference $ \left| {{\mathbf {v}} _ {n}} - {{\mathbf {v}} _ {s}} \right| $ should be of the order of $ \left(\hbar / Mb \right) \ln \left( b / \xi \right) $, where $ \xi $ is the radius of the vortex core \cite {bib:Vin2}. The transition to the turbulent state T-2 should occur when the normal component reaches the critical velocity $ {{R} _ {n}} {{\eta} _ {n}} / {{\rho} _ {n}} b $ ($ {{R} _ {n}} $ is the Reynolds number). The existence of critical velocities determines the maximum allowable size $ b $, for which the laminar motion of the fluid is preserved. This imposes a restriction on the electric potential (\ref {eq:22}).

Let us present the restrictions on the magnitude of the electric potential in the laminar flow regime. For $ \gamma \sim1 $ this restriction is connected with the fulfillment of the condition $ \left | {{\mathbf {v}} _ {n}} - {{\mathbf {v}} _ {s}} \right| <\left(\hbar / Mb \right) \ln \left(b / \xi \right) $
\begin {align}
\varphi <\frac {4 \pi \hbar} {3M} \frac {\alpha H} {Mc} \frac {{{\rho} _ {s}} {{\rho} _ {n}}} { \rho} \ln \left (\frac {b} {\xi} \right) \equiv {{\varphi} _ {c1}}. \label {eq:23}
\end {align}
For $ \gamma \ll 1 $, the restriction is imposed on the speed of the normal component $ \left \langle {{v} _ {n}} \right \rangle <{{R} _ {n}} {{\eta} _ {n }} / {{\rho} _ {n}} b $. It follows that
\begin {align}
\varphi <\frac {4 \pi} {3} \frac {\alpha H} {Mc} {{R} _ {n}} {{\eta} _ {n}} \equiv {{\varphi} _ {c2}}. \label {eq: 24}
\end {align}
Considering the temperature range in which $ {{\rho} _ {s}} {\sim} {{\rho} _ {n}} \sim \rho $, we find that $ {{\varphi} _ { c2}}> {{\varphi} _ {c1}} $. So for $ H = 10 $\,T, $ \alpha = 2\cdot{{10} ^ {- 25}} $\,cm$^3$, $ {{R} _ {n}} = 2 \cdot {{10} ^ {3}} $, $ {{\eta} _ {n}} = 2 $\,$\mu$P and $ \rho = {{10} ^ {- 1}} $\,g/cm$^3 $ we have $ {{\varphi} _ {c1}} = 2 \cdot {{10} ^ {- 8}} $\,V and $ {{\varphi} _ {c2}} = 5 \cdot {{10} ^ {- 7}} $\,V. Thus, to obtain the highest potential $ \varphi $, it is necessary to use not a circular capillary, but a slit, and the magnetic field should be directed across the slit. Now we can find the transverse gap size $ {{b} _ {c}} $ corresponding to the largest potential $ {{\varphi} _ {c2}} $. For a capillary of length $ L = 1 $\,cm, between the ends of which a temperature difference of $ \Delta T = {{10} ^ {- 3}} $\,K is created, from (\ref{eq:22}), in the specified temperature range, we find $ {{b} _ {c}} = {{10} ^ {- 3}} $\,cm.

Up to this point it was supposed that the temperature difference between the ends of the capillary does not depend on time. However, the effect is also possible in the case when such a difference periodically depends on time, i.e. in the condition of propagation of second sound in He\nobreakdash-II.

Let He\nobreakdash-II fill the cylinder of a circular cross-section of radius $ R $ and the second sound propagates along the axis of the cylinder $ z $. It is known that, in neglecting thermal expansion, in the first sound waves, pressure and density oscillations occur, and only temperature and entropy oscillate in the second sound waves. When thermal expansion is taken into account, there is a bond between the oscillations of the second and first sounds, and as a result, when the second sound propagates, pressure oscillations occur. These oscillations lead to a nonzero mass flow $ {{j} _ {z}} $, which is associated with the oscillating part of the temperature $ T'$. In the case of $ R \gg \lambda $, where $ \lambda $ is the wavelength of the second sound, the dependence of $ {{j} _ {z}} $ on the radial coordinate $ r $ can be neglected and the flow $ {{j} _ { z}} $ is related to the $ T'$ relation (see, for example, \cite{bib:Kh})
\begin{align}
{{j}_{z}}=-\frac{\beta \rho {{u}_{1}}^{2}{{u}_{2}}}{{{u}_{1}}^{2}-{{u}_{2}}^{2}}T',\label{eq:25}
\end{align}
where $ {{u} _ {1}} $ and $ {{u} _ {2}} $ are the speeds of the first and second sounds, respectively, $ \beta $ is the coefficient of thermal expansion. If the external magnetic field is directed across the cylinder (along the $ y $ axis), then in helium, according to (\ref{eq:3}), electric polarization  $ {{P} _ {x}} = - \alpha H {{ j} _ {z}} / Mc $ occurs. The electric potential outside the cylinder is obtained by substituting this polarization into (\ref{eq:9}). Details of the calculation of the arising integral can be found in \cite{bib:SK2}. Considering that for $ T> 0.8$\,K the condition $ {{u} _ {1}} \gg {{u} _ {2}} $ is fulfilled and assuming that the height of the cylinder $ L $ substantially exceeds the wavelength of the second sound $ \lambda $, we find the expression for the electric potential at the point with coordinates $ \left({{r} _ {0}}, {{\theta} _ {0}}, 0 \right) $
\begin{align}
\varphi \left( {{r}_{0}},{{\theta }_{0}},t \right)=4\pi \beta \rho {{u}_{2}}\frac{\alpha HR}{Mc}{{I}_{1}}(kR){{K}_{1}}\left( k{{r}_{0}} \right)\cdot{{T}_{a}}'\cos {{\theta }_{0}} \cos \left( {{u}_{2}}kt \right).\label{eq:26}
\end{align}
Here $ k $ is the wave number, $ {{T} _ {a}}'$ is the amplitude of the temperature in the second sound wave, $ {{I} _ {1}} $ and $ {{K} _ {1}} $ are first order modified Bessel functions of the first and second kind, respectively.

Above it was assumed that $ R \gg \lambda $ and we neglected the dependence of the flow $ {{j} _ {z}} $ on the radial coordinate $ r $. However, if the cylinder radius is of the order of the second sound wavelength, such an assumption is unjustified. In this case, to solve the problem of the propagation of a second sound correctly, one should use the complete system of hydrodynamic equations for a superfluid liquid, supplemented by boundary conditions. It is convenient to use the approach developed in \cite{bib:AdamOld}. If the conditions  $ {{u} _ {1}} \gg {{u} _ {2}} $ and $ \lambda \gg R \gg {{\lambda} _ {\eta}} $ are satisfied, where $ {{\lambda} _ {\eta}} = \sqrt {2 \pi \eta_n / \nu {{\rho} _ {n}}} $ is the length of a viscous wave ($ \nu $ is the frequency of the second sound), we get for the mass flow
\begin{align}
{{j}_{z}}=-\left( \beta \rho {{u}_{2}}+\frac{{{\rho }_{s}}S}{{{u}_{2}}}\frac{{{J}_{0}}\left( {{k}_{\eta }}r \right)}{{{J}_{0}}\left( {{k}_{\eta }}R \right)} \right)T',\label{eq:28}
\end{align}
where $ {{J} _ {0}} $ is the zero order Bessel function of the first kind, $ {{k} _ {\eta}} $ is the wave number of a viscous wave. Using (\ref{eq:3}) and (\ref{eq:28}) we find from (\ref{eq:9}) the expression for the electric potential at the point with coordinates $ \left({{r} _ { 0}}, {{\theta} _ {0}}, 0 \right) $
\begin{multline}
\varphi \left( {{x}_{0}},{{\theta }_{0}},t \right)=4\pi \frac{\alpha HR}{Mc}{{K}_{1}}\left( k{{r}_{0}} \right)\left[ \beta \rho {{u}_{2}}{{I}_{1}}(kR)-\right.\\\left.-\frac{{{\rho }_{s}}S}{{{u}_{2}}}\frac{{{\lambda }_{\eta }}}{\lambda }\cos \left( \frac{\pi }{8} \right) \right]{{T}_{a}}'\cos {{\theta }_{0}} \cos \left( {{u}_{2}}kt \right).\label{eq:29}
\end{multline}
This potential, like the potential in (\ref{eq:26}), depends on the polar angle of the observation point according to the cosine law and vanishes for $ {{\theta} _ {0}} = \pi / 2 $. In other words, on a plane parallel to the magnetic field $ \mathbf {H} $ and passing through the axis of the cylinder, the potential is equal to zero.

Note that the expressions for the potential $ \varphi $ during the propagation of the second sound, as in the static case, are obtained under the assumption of a laminar regime. Unfortunately, we are not aware of the analytical criteria for the transition from a laminar to a turbulent regime in the case of the propagation of second sound. In experiments, the radius of a cylindrical sample usually lies in the range $ R = 0.1...1 $\,cm  (see, for example, \cite{bib:Tough}). In this case, the amplitude value of the temperature corresponding to the laminar region does not exceed several millikelvin. For $ T = 1.5 $\,K, $ H = 10 $\,T, $ \nu = 400 $\,Hz, $ {{T} _ {a}} '= {{10} ^ {-3}} $\,K and $ R = 0.5 $\,cm, the potential on the surface of the cylinder at the point $ {{r} _ {0}} = R $, $ {{\theta} _ {0} } = 0 $ equals $ \varphi = 4 \cdot {{10} ^ {- 8}} $\,V.

In conclusion, we note that in experiments with a second sound, the electric fields predicted in this work should be distinguished from the electric fields reported in \cite{bib:R}. The potentials (\ref{eq:26}) and (\ref{eq:29}) have a characteristic angular dependence. In addition, we are talking about electric fields outside the sample, while the potential difference in \cite{bib:R} is measured inside the sample.

\vfill\eject


\begin{thebibliography}{27}
\bibitem{bib:R}
A.\,S. Rybalko, Low Temp. Phys. {\bf 30}, 994 (2004).

\bibitem{bib:RR}
A.\,S. Rybalko and S.\,P. Rubets, Low Temp. Phys. {\bf 31}, 623 (2005).

\bibitem{bib:Cha1}
T.\,V. Chagovets, Low Temp. Phys. {\bf 42}, 176 (2016).

\bibitem{bib:Cha2}
T.\,V. Chagovets, Physics B {\bf 488}, 62 (2016).

\bibitem{bib:Cha3}
T.\,V. Chagovets, J. Low Temp. Phys. {\bf 187}, 383 (2017).

\bibitem{bib:Jap}
H. Yayama, Y. Nishimura, H. Uchiyama, H. Kawai, Jean-Paul van Woensel and A.\,G. Hafez, Low Temp. Phys. {\bf 44}, 1386 (2018).

\bibitem{bib:M}
L.\,A. Melnikovsky, J. Low Temp. Phys. {\bf 148}, 559 (2007).

\bibitem{bib:N}
V.\,D. Natsik, Low Temp. Phys. {\bf 31}, 915 (2005).

\bibitem{bib:SR}
S.\,I. Shebchenko and A.\,S. Rukin, Low Temp. Phys. {\bf 36}, 146 (2010).

\bibitem{bib:SK1}
S.\,I. Shevchenko and A.\,M. Konstantinov, J. Low Temp. Phys. {\bf 185}, 384 (2016).

\bibitem{bib:AN1}
I.\,N. Adamenko and E.\,K. Nemchenko, Low Temp. Phys. {\bf 41}, 495 (2015).

\bibitem{bib:AN2}
I.\,N. Adamenko and E.\,K. Nemchenko, Low Temp. Phys. {\bf 42}, 258 (2016).

\bibitem{bib:SRJetpLett}
S.\,I. Shevchenko and A.\,S. Rukin, JETP Lett. {\bf 90}, 42 (2009).

\bibitem{bib:G}
E.\,D. Gutliansky, Low Temp. Phys. {\bf 35}, 748 (2009).

\bibitem{bib:Tom}
M.\,D. Tomchenko, Phys. Rev. B {\bf 83}, 094512 (2011).

\bibitem{bib:SKJetpLett}
S.\,I. Shevchenko and A.\,M. Konstantinov, JETP Lett. {\bf 104}, 489 (2016).

\bibitem{bib:SK2}
S.\,I. Shevchenko and A.\,M. Konstantinov, J. Low Temp. Phys. {\bf 194}, 1 (2019).

\bibitem{bib:Keesom}
W.\,H. Keesom, A.\,P. Keesom and B.\,F. Saris, Physica {\bf 5}, 281 (1938).

\bibitem{bib:LL8}
L.\,D. Landau, E.\,M. Lifshitz, \emph{Electrodynamics of Continuous Media}, (Pergamon Press 1960).

\bibitem{bib:LL6}
L.\,D. Landau, E.\,M. Lifshitz, \emph{Fluid Mechanics}, (Butterworth-Heinemann 1987).

\bibitem{bib:Vin1}
W.\,F. Vinen, Proc. R. Soc. Lond. A {\bf 240}, 114 (1957).

\bibitem{bib:TurbPRL}
D.\,J. Melotte and C.\,F. Barenghi, Phys. Rev. Lett. {\bf 80}, 4181 (1998).

\bibitem{bib:Tsu}
S. Yui and M. Tsubota, J. Phys., Conf. Ser. {\bf 568}, 012028 (2014).

\bibitem{bib:Tough}
J.\,T. Tough, Progress in Low Temp. Phys. {\bf 8}, 133 (1982).

\bibitem{bib:Vin2}
W.\,F. Vinen and J.\,J. Niemela, J. Low Temp. Phys. {\bf 128}, 167 (2002).

\bibitem{bib:Kh}
I.\,M. Khalatnikov, \emph{An introduction to th Theory of Superfluidity}, (Perseus Publishing, Cambrige, 2000).

\bibitem{bib:AdamOld}
I.\,N. Adamenko and M.\,I. Kaganov, JETP {\bf 26}, 394 (1968).

\end{thebibliography}
\end{document}